\documentclass[twocolumn,showpacs,preprintnumbers,amsmath,amssymb,prl,superscriptaddress,longbibliography]{revtex4-1}
\usepackage{bbm}
\usepackage{mathrsfs}
\usepackage{graphicx}
\usepackage{dcolumn}
\usepackage{bm}
\usepackage{amsmath}
\usepackage{amsfonts}
\usepackage{color}
\usepackage[colorlinks=true,citecolor=blue,anchorcolor=cyan]{hyperref}
\usepackage{floatrow}

\begin{document}

\title{Moir\'e Engineering and Topological Flat Bands in Twisted Orbital-Active Bilayers}
\author{Huan Wang}
\affiliation{State Key Laboratory of Surface Physics and Department of Physics, Fudan University, Shanghai 200433, China}
\author{Yadong Jiang}
\affiliation{State Key Laboratory of Surface Physics and Department of Physics, Fudan University, Shanghai 200433, China}
\author{Zhaochen Liu}
\affiliation{State Key Laboratory of Surface Physics and Department of Physics, Fudan University, Shanghai 200433, China}
\author{Jing Wang}
\thanks{wjingphys@fudan.edu.cn}
\affiliation{State Key Laboratory of Surface Physics and Department of Physics, Fudan University, Shanghai 200433, China}
\affiliation{Institute for Nanoelectronic Devices and Quantum Computing, Fudan University, Shanghai 200433, China}
\affiliation{Zhangjiang Fudan International Innovation Center, Fudan University, Shanghai 201210, China}

\begin{abstract}
Topological flat bands at the Fermi level offer a promising platform to study a variety of intriguing correlated phase of matter. Here we present band engineering in the twisted orbital-active bilayers with spin-orbit coupling. The symmetry constraints on the interlayer coupling that determines the effective potential for low-energy physics of moir\'e electrons are exhaustively derived for two-dimensional point groups. We find the line graph or biparticle sublattice of moir\'e pattern emerge with a minimal $C_3$ symmetry, which exhibit isolated electronic flat bands with nontrivial topology. The band flatness is insensitive to the twist angle since they come from the interference effect. Armed with this guiding principle, we predict that twisted bilayers of 2H-PbS$_2$ and CdS realize the salient physics to engineer two-dimensional topological quantum phases. At small twist angles, PbS$_2$ heterostructures give rise to an emergent moir\'e Kagom\'e lattice, while CdS heterostructures lead to an emergent moir\'e honeycomb lattice, and both of them host moir\'e quantum spin Hall insulators with almost flat topological bands. We further study superconductivity of these two systems with local attractive interactions. The superfluid weight and Berezinskii-Kosterlitz-Thouless temperature are determined by multiband processes and quantum geometry of the band in the flat-band limit when the pairing potential exceeds the band width. Our results demonstrate twisted bilayers with multi-orbitals as a promising tunable platform to realize correlated topological phases.

\end{abstract}

\date{\today}


\maketitle

\emph{Introduction.} 
Recently twisted van der Waals heterostructures are becoming a novel tunable platform to realize various intriguing electronic phases in two dimensions~\cite{andrei2020,balents2020,carr2020,kennes2021,andrei2021}. Two prime examples are twisted graphene and transition metal dichalcogenide (TMD) multilayers, where correlated insulation, superconductivity and interaction-induced quantum anomalous Hall effect have been discovered~\cite{bistritzer2011,cao2018a,cao2018b,yankowitz2019,lu2019,sharpe2019,chengr2019b,liuxm2020,serlin2020,stepanov2020,cao2020a,cao2020b,li2021,zhang2021,xie2022}. These interesting phenomena stem from the quenched kinetic energy due to quantum interference effect in the moir\'e superlattice, where strong electron interaction dominates in the almost dispersionless bands near or at the Fermi level~\cite{kang2019b,kerelsky2019,xieyl2019,lian2021}. The highly tunable twist angle and gate provide experimental access to study the paradigmatic models of strongly-correlated electrons~\cite{kennes2021}. However, the band width of the almost flat bands in twisted graphene multilayers varies rapidly with the twist angle, which poses the main experimental difficulty requiring the angle precision to be better than $0.1^\circ$ in order to achieve the band flatness~\cite{bistritzer2011}. Moreover, the superfluid weight is bounded when the flat band is topological~\cite{peotta2015,hu2019,julku2020,xie2020,herzog2022,torma2021,peri2021,verma2021}, which could increase the superconducting critical temperature. Heuristically, topological bands contain extended states with algebraically decaying tails~\cite{marzari2012}, which participate in the superconductivity~\cite{kopnin2011,qi2011,sato2017}. Therefore, it is important to find a topological moiré system, wherein the quasiflat bands have topological properties and its band width is insensitive to twist angle.

It is well known that the flat bands exist in the line and split graph of biparticle lattices~\cite{mielke1991a,mielke1991b,tasaki1998,wu2007,bergman2008,liu2014,ma2020,calugaru2022}. These bands have been exhaustively predicted in stoichiometric materials~\cite{regnault2022,liu2021}. However, the crystalline materials with isolated topological flat bands are still lacking. The previous studies have shown that new lattice structure and effective orbitals will emerge after twist in the moir\'e pattern, for example the $\Gamma$-valley TMD moir\'e bands~\cite{angeli2021,pei2022} and magnetic moir\'e surface bands in topological insulators~\cite{liu2022,lian2020}. There, a two-orbital model on a honeycomb lattice and a single-orbital model on a Kagom\'e lattice emerge from the $\Gamma$-valley moir\'e bands, which exhibit flat bands in their spectra. However, these flat bands are gapless due to absense of spin-orbit coupling (SOC). In this work, we present the topological moir\'e engineering in the twisted orbital-active bilayers with SOC, and predict isolated flat bands with nontrivial topology guaranteed by certain symmetries. We further propose two candidate materials 2H-PbS$_2$ and CdS to demonstrate the topological flat bands near or at the Fermi level, opening up an exotic regime for experimental and theoretical investigation. Finally, we study superconductivity in these systems with local attractive interactions.

\emph{Twisted multi-orbital bilayer.} We present the moir\'e engineering by starting from the twisted bilayer with generic multi-orbitals in each layer, which naturally exhibit atomic SOC. We consider stacking of two identical layers with the $z$ axis as a normal direction. The angle mismatch is achieved when rotating the upper layer and the lower layer by $+\varphi/2$ and $-\varphi/2$ around the $z$ axis, respectively. $\mathbf{a}_1$ and $\mathbf{a}_2$ are the Bravais unit vectors of each layer. Now the twist leads to the moir\'e pattern with periodicity of $\mathbf{L}_1$ and $\mathbf{L}_2$ as
\begin{equation}\label{vector}
\mathbf{L}_i=-\frac{\hat{z}\times\mathbf{a}_i}{2\sin(\varphi/2)}.
\end{equation}
The system is locally well approximated by an untwisted bilayer in a small-angle limit, while globally the moir\'e pattern is described by spatial modulation in the stacking condition. The generic effective model for such a system under the influence of the moir\'e pattern is written as
\begin{equation}\label{model}
H_{\text{eff}} = 
\begin{pmatrix} 
H_u(-i\boldsymbol\nabla) & V(\mathbf{r}) \\ 
V^\dagger(\mathbf{r}) & H_l(-i\boldsymbol\nabla) 
\end{pmatrix},
\end{equation}
where the diagonal terms $H_{u,l}(-i\boldsymbol\nabla)$ are the kinetic energy in upper and lower layer with the twist angle $\pm\varphi/2$, and the off-diagonal term $V(\mathbf{r})$ is the interlayer coupling, i.e., the moir\'e potential, which depends on the local stacking condition. The stacking is described by in-plane relative shift of the two layers, $\mathbf{t}(\mathbf{r})\equiv\mathbf{t}_u(\mathbf{r})-\mathbf{t}_l(\mathbf{r})$, where $\mathbf{t}_{u/l}(\mathbf{r})=\pm\sin(\varphi/2)\hat{z}\times\mathbf{r}$. The origin $\mathbf{t}=0$ corresponds to the upper layer right-on-top of the lower layer, i.e., vertical shift. $V(\mathbf{r})$ depends on position through the spatial dependence of $\mathbf{t}(\mathbf{r})$. In the small-angle limit, $H_{u/l}$ containing spatial derivative becomes less important and $V(\mathbf{r})$ dominates. As such, we treat $H_{u/l}$ as perturbation, and diagonalize $H_{\text{eff}}$ with only the off-diagonal components~\cite{jung2014,wu2019,kariyado2019}. We can see the diagonalized $V(\mathbf{r})$ plays a role of potential energy, and its extremum could trap band electrons and may form new lattice structure. Therefore knowing the properties of $V(\mathbf{r})$ is important to understand the electronic structure in twisted bilayers. 

Central to the present work, we consider the multi-orbitals in each layer and require each layer to have quadratic band dispersion, which is different from that in twisted graphene and TMD bilayer. Thus when SOC is neglected, the Fourier transformation of $H(-i\boldsymbol\nabla)$ is degenerate at the \emph{base momentum} $\mathbf{k}_0$ and lowest $\mathbf{k}\cdot\mathbf{p}$ expansion around it is $\delta\mathbf{k}^2$, i.e., quadratic band touching. The little group of $\mathbf{k}_0$ must be high symmetry. Table~\ref{tab1} lists all of the possible planar point groups with 2d irreducible representation for the spinless case, the corresponding base momentum and angular momentum $\ell_z$ of degenerate multi-orbitals. We assume that each layer has the full symmetry of that in Table~\ref{tab1}. Now the emergent moir\'e lattice is determined by $V(\mathbf{r})$, whose concrete form is constrained by the corresponding symmetry~\cite{balents2019}. Here we demonstrate $V(\mathbf{r})$ and emergent moir\'e lattice for the typical $C_{6v}$ here, and relegate the detailed calculation for the symmetry groups in Table~\ref{tab1} in Supplementary Materials.

\begin{table}[t]
\caption{The two-dimensional point groups with 2d irreducible representation for the spinless case under time-reversal symmetry. The angular momentum $\ell_z$ can be from atomic orbitals as well as the lattice.} 
\begin{center}\label{tab1}
\renewcommand{\arraystretch}{1.4}
\begin{tabular*}{3.4in}
{@{\extracolsep{\fill}}ccccccc}
\hline
\hline
symmetry group & $C_{3}$ & $C_{3v}$ & $C_{4}$ & $C_{4v}$ & $C_{6}$ & $C_{6v}$ 
\\
\hline
base momentum $\mathbf{k}_0$ & $\Gamma$ & $\Gamma$ & $\Gamma,M$ & $\Gamma,M$ & $\Gamma$ & $\Gamma$ 
\\
multi-orbitals $\ell_z$ & $\pm1$ & $\pm1$ & $\pm1$ & $\pm1$ & $\pm1$ or $\pm2$ & $\pm1$ or $\pm2$
\\
\hline
\hline
\end{tabular*}
\end{center}
\end{table}

\emph{Effective model and emergent lattice.} For the degenerate orbitals with $\ell_z=\pm1$, the $\mathbf{k}\cdot\mathbf{p}$ model of electronic state in each layer with $C_{6v}$ symmetry at $\Gamma$ point is 
\begin{eqnarray}\label{single}
H_{\text{single}}(\mathbf{k})&=&-\frac{\hbar^2}{2m^\star}\left\{(k_x^2+k_y^2)\hat{\mathbf{1}}+ \kappa\left[(k_x^2-k_y^2)\hat{\boldsymbol\tau}^x\right.\right.
\nonumber
\\
&&\left.\left.+ 2k_xk_y\hat{\boldsymbol\tau}^y\right]\right\}+\frac{\lambda_{\text{soc}}}{2}\hat{\boldsymbol\tau}^z\hat{\boldsymbol\sigma}^z,
\end{eqnarray}
where $\hat{\boldsymbol\tau}^i$ and $\hat{\boldsymbol\sigma}^i$ ($i=x,y,z$) are Pauli matrices acting on orbital and spin, respectively. $m^\star$ is the average effective mass, $\kappa=(m_+-m_-)/(m_++m_-)$ is real parameterizing the effective mass ratio of light ($m_-$) and heavy ($m_+$) bands at $\Gamma$. $\lambda_{\text{soc}}$ is the SOC strength, which lifts the orbital degeneracy and opens up a topological gap at $\Gamma$. $V(\mathbf{r})$ is spatially periodic with the periodicity $\mathbf{L}_1$ and $\mathbf{L}_2$, which can be Fourier expanded as
\begin{equation}\label{potential}
V(\mathbf{r})=V_0+\sum_{n,i} V(\mathbf{g}_{n,i})e^{i\mathbf{g}_{n,i}\cdot\mathbf{r}},
\end{equation}
where $\mathbf{g}_{n,i}$ denote the six moir\'e reciprocal lattice vectors $i=1,2,3,4,5,6$ (related via $C_6$ rotations) to the $n$-th moir\'e Brillouin zone, $\mathbf{g}_{n,4}=-\mathbf{g}_{n,1}$, $\mathbf{g}_{n,5}=-\mathbf{g}_{n,2}$, $\mathbf{g}_{n,6}=-\mathbf{g}_{n,3}$. The $h\in C_{6v}$ symmetry $V(\mathbf{r})=hV(h\mathbf{r})h^{-1}$ and time-reversal $\mathcal{T}$ symmetry $\mathcal{T}^\dag V(\mathbf{r})\mathcal{T}=V(\mathbf{r})$ lead to $V(\mathbf{g}_{n,1})=V(\mathbf{g}_{n,4})$, $V(\mathbf{g}_{n,2})=V(\mathbf{g}_{n,5})$, $V(\mathbf{g}_{n,3})=V(\mathbf{g}_{n,6})$, and
\begin{eqnarray}
V(\mathbf{g}_{n,1}) &=& 
\begin{pmatrix} 
\alpha_n & \beta_n e^{-i\frac{\pi}{3}} \\ 
\beta_n e^{i\frac{\pi}{3}} & \alpha_n
\end{pmatrix},
V(\mathbf{g}_{n,2})= 
\begin{pmatrix} \alpha_n & -\beta_n\\ 
-\beta_n & \alpha_n 
\end{pmatrix},
\nonumber
\\
V(\mathbf{g}_{n,3}) &=&
\begin{pmatrix} 
\alpha_n & \beta_n e^{i\frac{\pi}{3}}\\ 
\beta_n e^{-i\frac{\pi}{3}} & \alpha_n 
\end{pmatrix}.
\end{eqnarray}
Here $\alpha_n$ and $\beta_n$ parameterizes the Fourier modes of interlayer coupling, and we expect the tunneling decays for higher harmonics~\cite{bistritzer2011}.

\begin{figure}[t]
\begin{center}
\includegraphics[width=3.4in,clip=true]{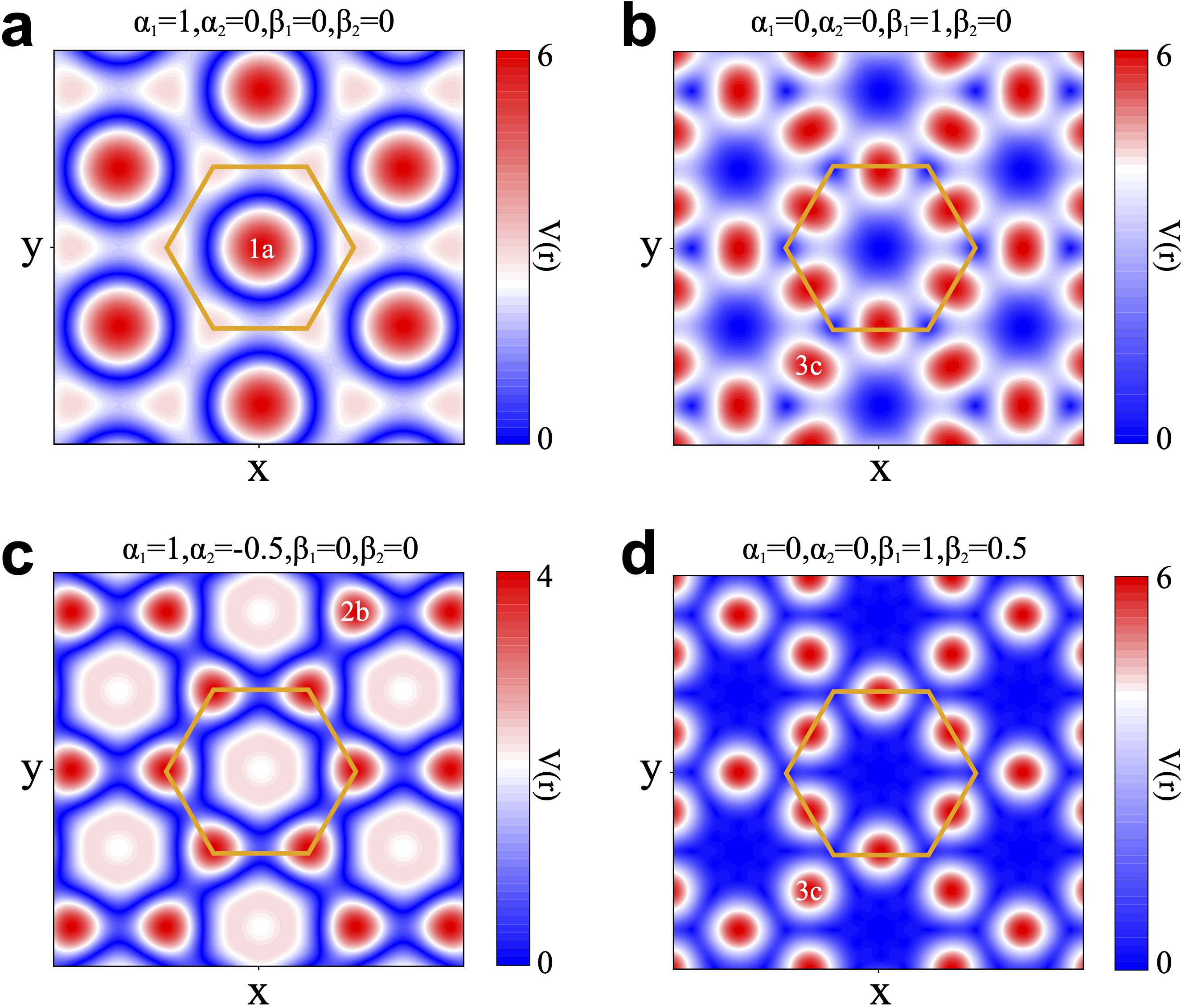}
\end{center} 
\caption{\textbf{Typical effective potentials of twisted bilayer with $C_{6v}$ symmetry}. The relative magnitudes between $\alpha_{1,2}$ and $\beta_{1,2}$ gives rise to different moir\'e lattice structure at the potential extremum. The maximal $1a$, $2b$, and $3c$ Wyckoff positions of a hexagonal lattice are labelled. For example,  \textbf{b}, \textbf{d}, Kagom\'e lattice from the Wyckoff position $3c$ emerges when $\beta_{1,2}$ dominates. }
\label{fig1}
\end{figure}

The emergent lattice structure of moir\'e pattern is explicitly seen by diagonalizing $H_{\text{eff}}$ with only $V(\mathbf{r})$ keeping up to second order terms. The parameters $\alpha_{1,2}$ and $\beta_{1,2}$ for realistic materials can be obtained by fitting the band structure from first-principles calculations.  Fig.~\ref{fig1} demonstrates the \emph{typical} effective potential of twisted bilayer with $C_{6v}$ symmetry. For example, when $\alpha_{1,2}$ dominate, one get the triangle or honeycomb lattices; when $\beta_{1,2}$ dominates, one get the Kagom\'e or ruby lattices. As such, the line graph of biparticle lattice emerges and the topological flat band is expected~\cite{wang2021}. The effective lattices for twisted bilayer with symmetry groups in Table~\ref{tab1} are exhaustively calculated in Supplementary Materials. Here we focus on those lattices which enforces the topological flat bands and draw the conclusion: for twisted orbital-active hexagonal bilayer ($C_3$,$C_{3v}$,$C_6$,$C_{6v}$), honeycomb and Kagom\'e lattices could emerge, where isolated topological quasiflat bands are obtained in the presence of SOC; while for twisted tetragonal bilayer ($C_4$,$C_{4v}$), square and Lieb (as well as line graph of Lieb) lattices could emerge, where the latter is fine tuned, thus flat bands are not guaranteed. Next we predict two candidate materials to realize the salient physics described here. 

\begin{figure*} 
\begin{center}
\includegraphics[width=6.5in,clip=true]{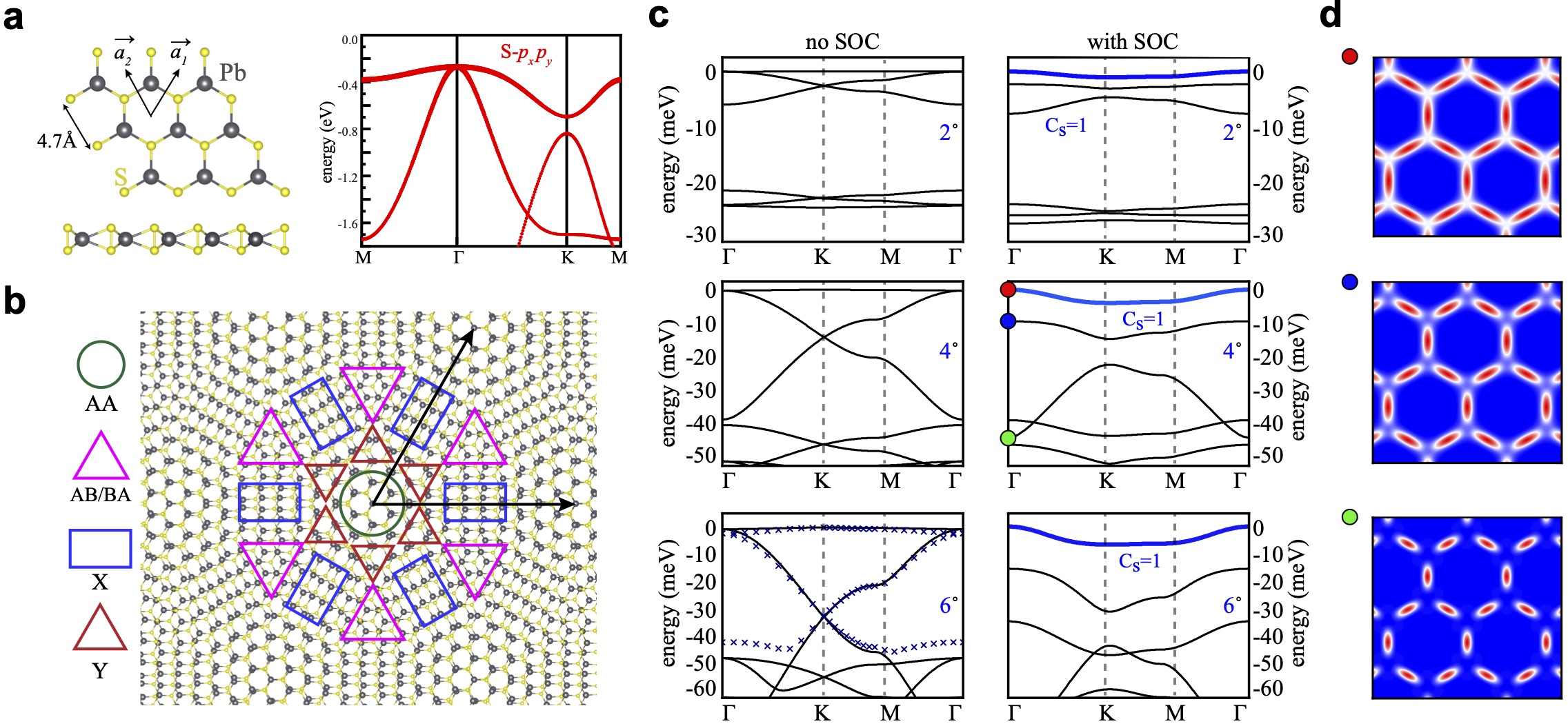}
\end{center} 
\caption{\textbf{Moir\'e pattern and topological flat bands in twisted PbS$_2$ bilayer}. \textbf{a}, Crystal structure and electronic structure without SOC of monolayer 2H-PbS$_2$, the in-plane lattice vectors $\mathbf{a}_1$ and $\mathbf{a}_2$ are shown. \textbf{b}, In the small angle twisted PbS$_2$ bilayer, the interlayer stacking has AA, AB/BA, X and Y in the moir\'e unit cell, where $\mathbf{t}_{\text{AA}}(\mathbf{r})=0$, $\mathbf{t}_{\text{AB/BA}}(\mathbf{r})=\pm(\mathbf{a}_1+\mathbf{a}_2)/3$, $\mathbf{t}_{\text{X}}(\mathbf{r})=(\mathbf{a}_1+\mathbf{a}_2)/2$, and $\mathbf{t}_{\text{Y}}(\mathbf{r})=(\mathbf{a}_1+\mathbf{a}_2)/6$. \textbf{c}, The band structure of twisted bilayers from continuum model for three representative twist angles without (left column) and with (right column) SOC. The top-most valence bands demonstrate the flat bands on the emergent Kagom\'e lattice. The SOC lifts the quadratic band touching of Kagom\'e bands at $\Gamma$ and leads to isolated topological quasiflat bands (blue lines) with spin Chern number $C_s=\pm1$. For twist angle $6^\circ$ without SOC, the band structure from continuum model matches well with that from first-principles calculations (navy cross). \textbf{d}, The wave function density distribution of the three selected Bloch states that are circled in \textbf{c}, which clearly shows the emergent Kagom\'e lattice.}
\label{fig2}
\end{figure*}

\emph{Materials.} PbS$_2$ is an exfoliable semiconductor, and its monolayer with 2H structure has been introduced in Computational 2D Materials Database (C2DB)~\cite{C2DB_1,C2DB_2}. 2H-PbS$_2$ phase is more stable (total energy $0.47$~eV/u.c. lower) than its 1T phase. The band structure of monolayer 2H-PbS$_2$ without SOC from first-principles calculations is shown in Fig.~\ref{fig2}a. In contrast to group-VI TMD such as MoS$_2$ with 2H structure, the valence band maximum in 2H-PbS$_2$ is located at $\Gamma$, and without SOC it is characterized by two-fold degenerate chalcogen $p_x$, $p_y$ orbitals enforced by space group $P$-$6m2$. The atomic SOC lifts the degeneracy and introduces a gap of about $65$~meV. The parameters in Eq.~(\ref{single}) are obtained from first-principles calculations $m^\star= 0.7 m_0$, with $m_0$ the mass of free electron, $\kappa=0.9$, and $\lambda_{\text{soc}}=65$~meV. The parameters in $V(\mathbf{r})$ up to second order are obtained by fitting the \emph{ab initio} band structures with different stacking configurations such as AA, AB, X and Y shown in Fig.~\ref{fig2}b, $\alpha_1=-24.3$~meV, $\alpha_2=5.5$~meV, $\beta_1=26.8$~meV, and $\beta_2=10.3$~meV. The band structure from the continuum model and DFT calculations for twisted bilayer of $6^\circ$ without SOC matches well with each other in Fig.~\ref{fig2}c, justifying the validity of the fitted parameters.

The band structures for valence electrons calculated in Fig.~\ref{fig2}c show clearly the emergent Kagom\'e flat bands. To reveal the nature of moir\'e band physics, we identify the symmetries and centers of the Wannier orbitals underlying the moiré bands. The wave function density distribution of the three selected Bloch states that are circled in Fig.~\ref{fig2}c is generated by orbitals centered on a Kagom\'e lattice shown in Fig.~\ref{fig2}d. The SOC lifts the degeneracy at $\Gamma$, and the flat band is gapped from other dispersive band which becomes quasiflat and is topological with spin Chern number $C_s=\pm1$. The flatness of the top-most topological band is insensitive to the twist angle since they come from the interference effect.

\begin{figure*}
\begin{center}
\includegraphics[width=6.5in,clip=true]{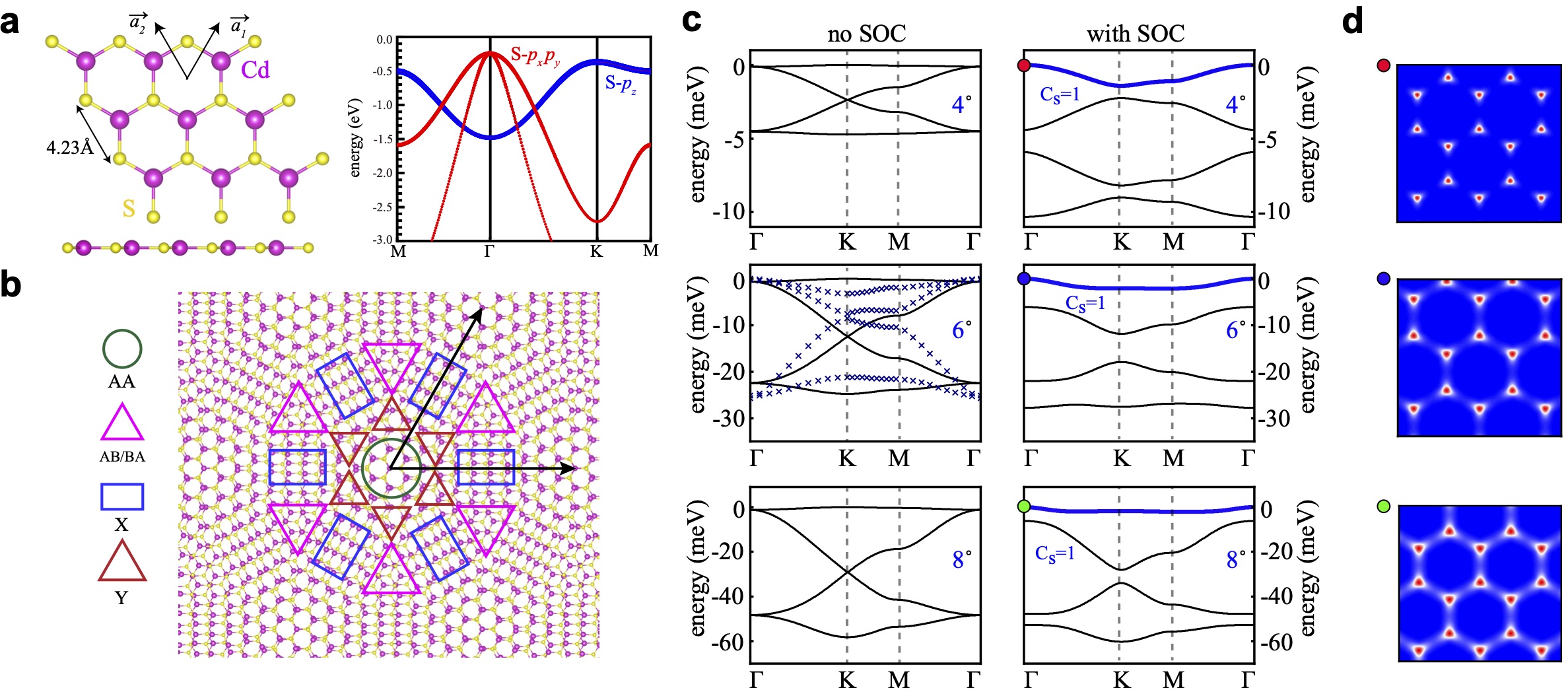}
\end{center} 
\caption{\textbf{Moir\'e pattern and topological flat bands in twisted CdS bilayer}. \textbf{a}, Top and side view of crystal structure of monolayer CdS, where the in-plane lattice vectors $\mathbf{a}_1$ and $\mathbf{a}_2$ are shown. The electronic structure without SOC from DFT calculation.  \textbf{b}, In the small angle twisted CdS bilayer, the interlayer stacking has AA, AB/BA, X and Y in the moir\'e unit cell. Here $\mathbf{t}_{\text{AA}}(\mathbf{r})=0$, $\mathbf{t}_{\text{AB/BA}}(\mathbf{r})=\pm(\mathbf{a}_1+\mathbf{a}_2)/3$, $\mathbf{t}_{\text{X}}(\mathbf{r})=(\mathbf{a}_1+\mathbf{a}_2)/2$, and $\mathbf{t}_{\text{Y}}(\mathbf{r})=(\mathbf{a}_1+\mathbf{a}_2)/6$. \textbf{c}, The band structure of twisted bilayers from continuum model for three representative twist angles without (left column) and with (right column) SOC. The top-most valence bands demonstrate the flat bands on the emergent honeycomb lattice. The SOC lifts the quadratic band touching of honeycomb bands at $\Gamma$ and leads to isolated topological quasiflat bands (blue lines) with spin Chern number $C_s=\pm1$. For twist angle $6^\circ$ without SOC, the band structure from continuum model qualitatively matches well with that from DFT calculations (navy cross). \textbf{d}, The wave function density distribution of the three selected Bloch states that are circled in \textbf{c}, which clearly shows the emergent honeycomb lattice.}
\label{fig3}
\end{figure*}

CdS is another exfoliable semiconductor, and its monolayer has been introduced in 2D Materials Encyclopedia~\cite{2dmatpedia}. The band structure of monolayer CdS without SOC from DFT calculations is shown in Fig.~\ref{fig3}a. Similar to monolayer 2H-PbS$_2$, the valence band maximum of monolayer CdS is located at $\Gamma$, which is composed of two-fold degenerate chalcogen $p_x$, $p_y$ orbitals enforced by space group $P3m1$. The atomic SOC lifts the degeneracy and introduces a gap of about $14$~meV. The parameters for CdS in Eq.~(\ref{single}) are obtained from DFT calculations $m^\star= 0.9m_0$, $\kappa=0.9$ and $\lambda_{\text{soc}}=13.4$~meV. The parameters in $V(\mathbf{r})$ are fitted with the \emph{ab initio} band structures as $\alpha_1=-45.6$~meV, $\alpha_2=41.5$~meV, $\beta_1=16.5$~meV, and $\beta_2=7$~meV.
 
The band structures for valence electron calculated in Fig.~\ref{fig3}c show the emergent honeycomb flat bands with degenerate $p_x$, $p_y$ orbitals near the Fermi level by employing topological quantum chemistry~\cite{bradlyn2017,slager2017}. The charge density distribution of the selected Bloch states circled in Fig.~\ref{fig3}c is generated by multiorbitals centered on a honeycomb lattice shown in Fig.~\ref{fig3}d. Such emergent honeycomb lattice can be explained by the maximum of $V(\mathbf{r})$ dominated by $\alpha_{1,2}$. The SOC lifts the degeneracy at $\Gamma$ and gap the flat band with spin Chern number $C_s=\pm1$. The salient low-energy physics of orbital-active biparticle lattice is quite unusual, since it is not energetically favorable in 2D crystalline materials for $p_x$, $p_y$ orbitals such as graphene due to $sp^2$ configuration. It is worth mentioning that the band structure for twist angle $4^\circ$ looks like the Kane-Mele type~\cite{kane2005b}, because now $\lambda_{\text{soc}}$ becomes the dominate energy scale and splits the degenerate $p_x$, $p_y$ orbitals into $p_\pm=p_x\pm ip_y$, as such the low-energy physics is characterized by single orbital.

\begin{figure}[htbp]  
\begin{center}
\includegraphics[width=3.4in,clip=true]{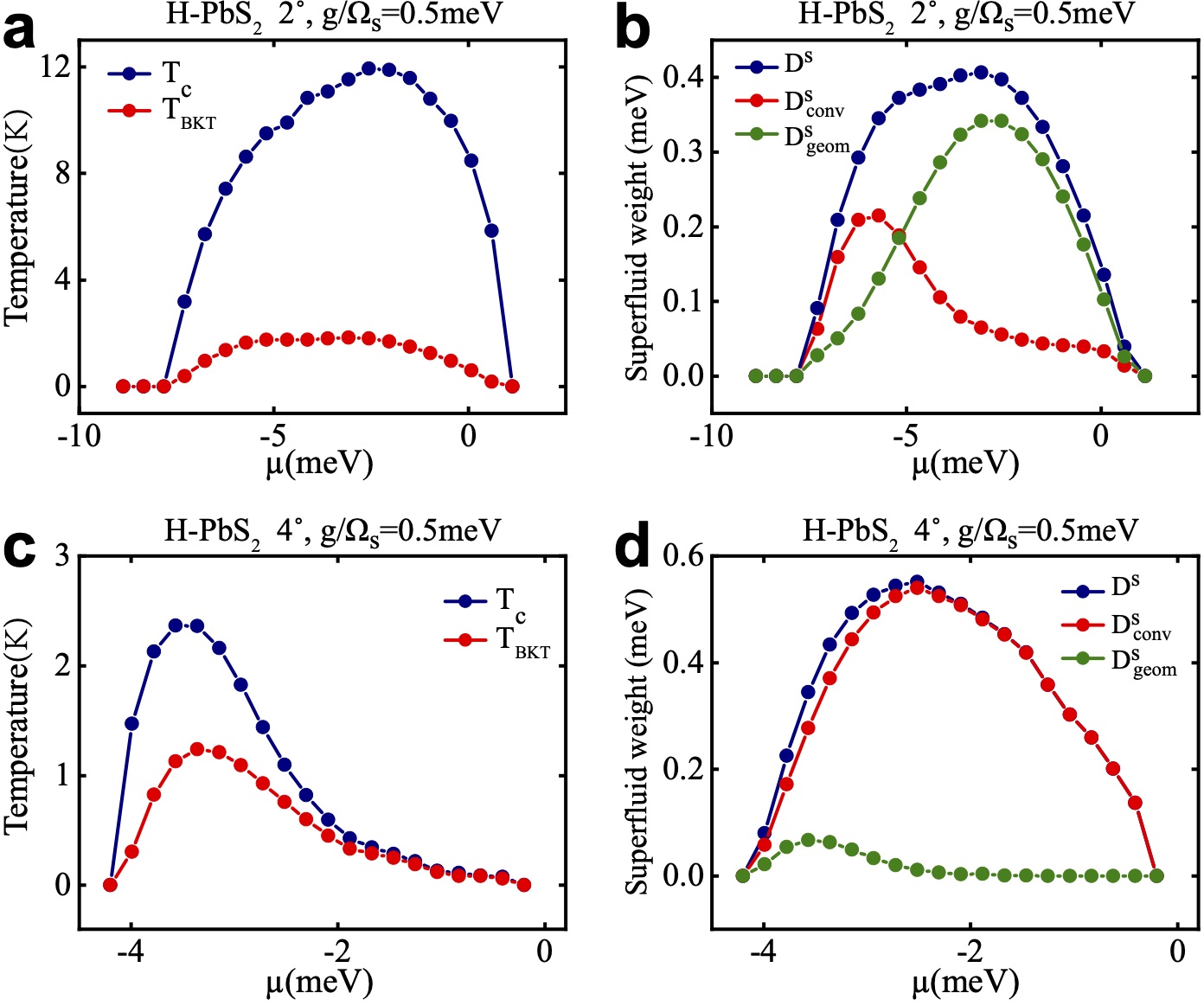}
\end{center} 
\caption{\textbf{Superfluid weight and BKT temperature.} \textbf{a},\textbf{c}, The critical temperature and BKT temperature as a function of chemical potential for Kagom\'e bands in twisted PbS$_2$ bilayer of $2^{\circ}$ and $4^{\circ}$, respectively. \textbf{b},\textbf{d}, The corresponding various components of superfluid weight.}
\label{fig4}
\end{figure}

\emph{Superfluid weight and BKT transition.} The tunable realization of isolated topological flat bands in twisted orbital-active bilayer provides an ideal platform for hosting correlated topological phases, including fractional quantum anomalous Hall effect~\cite{levin2009,maciejko2010,qi2011f,tang2011,sun2011,neupert2011,claassen2022} and unconventional superconductivity~\cite{balents2020,cao2018b}. Here we study superconductivity of twisted bilayer PbS$_2$ and CdS with local attractive interactions. We obtain the superfluid weight and Berezinskii-Kosterlitz-Thouless (BKT) transition temperature for low-energy continuum and tight-binding models.

We consider the local attractive interaction $H_{\text{int}}=-g\int dr \psi^{\dagger}_{l\tau\uparrow}\psi^{\dagger}_{l^\prime\tau^\prime\downarrow}\psi_{l^\prime\tau^\prime\downarrow}\psi_{l\tau\uparrow}$, where $l$ denotes layer and $\tau$ denotes orbital degrees of freedom. The local interaction has been used to study $s$-wave superconductivity, mediated by electron-phonon interaction in twisted bilayer graphene~\cite{wu2018a,lian2019}. Now the total Hamiltonian is $\mathcal{H}=H_{\text{eff}}-\mu N+H_{\text{int}}$, where $N$ is particle number operator and $\mu$ is the chemical potential. To determine the superfluid weight $D^s$, we first solve the order parameters from the BCS gap equation. After projecting the interaction term onto bands near the Fermi surface by $\psi_{l\tau\sigma}(\mathbf{k}+\mathbf{G})=\sum_{n} u_{n\sigma\mathbf{G}}(l,\tau,\mathbf{k}) c_{n\sigma}(\mathbf{k})$, where $u_{n\sigma}(\mathbf{k})$ is the Bloch wave function and $c_{n\sigma}(\mathbf{k})$ is the annihilation operator with band index $n$, spin $\sigma$ and reciprocal vector $\mathbf{G}$, the interacting Hamiltonian is solved by mean field approximation with order parameter matrix as $\Delta(\mathbf{k})_{n_3,n_4}=(1/A)\sum_{\mathbf{k}_1}[V_{\mathbf{k},-\mathbf{k},\mathbf{k}_1-\mathbf{k}}]^{n_1,n_4}_{n_2,n_3}\langle c^{\dagger}_{n_1,\uparrow}(\mathbf{k}_1)c^{\dagger}_{n_2,\downarrow}(-\mathbf{k}_1)\rangle$.  We solve the linearized gap equation to obtain the leading pairing instability and the critical temperature $T_c$ as 
\begin{equation}
\Delta^{\dagger}(\mathbf{k}_1)_{n_1,n_2}=\frac{1}{A}\sum_{k}[V_{\mathbf{k},-\mathbf{k},\mathbf{k}_1-\mathbf{k}}]^{n_1,n_4}_{n_2,n_3}f_{F,n_4,n_3}(\mathbf{k})\Delta^{\dagger}(\mathbf{k})_{n_4,n_3},
\end{equation}
where $f_{F,n_1,n_2}=\frac{1-n_F(\varepsilon_{n_1,\uparrow})-n_F(\varepsilon_{n_2,\downarrow})}{\varepsilon_{n_1,\uparrow}+\varepsilon_{n_2,\downarrow}}$ and $[V_{\mathbf{k},\mathbf{k}_1,\mathbf{q}}]^{n_1,n_4}_{n_2,n_3}=g\sum_{\mathbf{G}}\Lambda^{n_1,n_4}_{\mathbf{q}+\mathbf{G},\uparrow}(\mathbf{k})\Lambda^{n_2,n_3}_{-\mathbf{q}-\mathbf{G},\downarrow}(\mathbf{k}_1)$, $A$ is area of the sample, $n_F$ is the Fermi-Dirac distribution and $\varepsilon_{n_i,\sigma}$ is the band dispersion, $\Lambda^{n_1,n_2}_{\mathbf{q}+\mathbf{G},\sigma}(\mathbf{k})=\langle u_{n_1\sigma},\mathbf{k}+\mathbf{q}+\mathbf{G}|u_{n_2\sigma},\mathbf{k}\rangle$. The maximum of order parameter depends almost linearly on $g$, which is typical for generic flat band systems. Here we choose $g/\Omega_s=0.5$~meV based on previous studies on electron-phonon coupling in moir\'e system~\cite{wu2018a,lian2019}, where $\Omega_s$ is the area of the moir\'e unit cell. The accurate strength relies on microscopic details of electron-phonon coupling, which is beyond the scope of this paper.

As the system is two-dimensional, the transition to superconductivity is bound to occur at the BKT temperature $T_{\text{BKT}}$~\cite{berezinskii1971,kosterlitz1972,kosterlitz1973}, which can be determined from $k_BT_{\text{BKT}}=(\pi/8)\sqrt{\text{det}(D^s(T_{\text{BKT}}))}$. From the linear response theory, $D^s$ is the zero-frequency, long-wavelength limit of the current-current response function $D^s_{\mu\nu}=\lim_{\mathbf{q}\rightarrow0}[\lim_{\omega\rightarrow0}K_{\mu\nu}(\mathbf{q},\omega)]$~\cite{liang2017}, where $\mu,\nu\in(x,y)$. The local attractive interaction, yielding the spin-singlet $s$-wave pairing, conserves the underlying $C_3$ rotational symmetry of the twisted bilayer lattices which enforces $D^s$ to be isotropic, i.e., $D^s_{xx}=D^s_{yy}$ and $D^s_{xy}=D^s_{yx}=0$. Fig.~\ref{fig4}c and~\ref{fig4}d depict the numerical results of $T_{\text{BKT}}$ and $D^s$ for twisted PbS$_2$ bilayer at twisting angle $\theta=4^{\circ}$.  Since the band width and interband gaps are greater than the interaction strength, we could only keep the Kagom\'e bands. As we can see, $T_{c}$ follows the density of state as in the BCS superconductor, and the superfluid weight is dominated by the conventional part $D^s_{\text{conv}}$ which is proportional to the group velocity of electronic bands around the Fermi level. $D^s=D^s_{\text{conv}}+D^s_{\text{geom}}$, where $D^s_{\text{geom}}$ is a multiband effect depending on the overlap of the Bloch states and their momentum derivatives. When the twist angle further decreases to $\theta=2^{\circ}$, the band widths become much smaller as shown in Fig.~\ref{fig2}c. The reduced band width indicates the more localized wannier orbitals in real space, thus the Hamiltonian reduces to Kagom\'e lattice attractive Hubbard model with Hubbard $\left|U\right|$ comparable with the band width. As depicted in Fig.~\ref{fig4}a and~\ref{fig4}b, the increased $|U|$ and density of state are reflected in relatively large $T_c$, and large proportion of the superfluid weight comes from geometric part when the chemical potential enters flat band. All these features were verified for the lattice model calculation in Supplementary Materials.

\emph{Discussion.} The moir\'e engineering of twisted orbital-active bilayer leads to interesting and unique topological electronic structure. Besides the superconductivity, the realization of isolated topological flat bands here further provides a promising and tunable platform to realize a variety of correlated topological states of matter, including fractional topological insulator and fractional quantum anomalous Hall effect. The guiding principle for searching realistic materials is to find hexagonal lattice with minimal $C_3$ symmetry, where the low-energy physics is at $\Gamma$. Therefore stronger band flattening at larger twist angles in the quadratic cases is expected comparing with the linear cases. All of these requirements are quite common for valence electrons in semiconductors, for example, PbSe$_2$, SnS$_2$, SnSe$_2$, SnTe$_2$, CdSe, CdTe, ZnSe, etc~\cite{C2DB_1,C2DB_2}. Interestingly, the emergent Kagom\'e lattice could be extended to magnetic moir\'e systems, which is left for future work.

\begin{acknowledgments}
This work is supported by the National Key Research Program of China under Grant No.~2019YFA0308404, the Natural Science Foundation of China through Grant No.~12174066, Science and Technology Commission of Shanghai Municipality under Grant No.~20JC1415900, the Innovation Program for Quantum Science and Technology through Grant No.~2021ZD0302600, Shanghai Municipal Science and Technology Major Project under Grant No.~2019SHZDZX01. H.W. and Y.J. contributed equally to this work.
\end{acknowledgments}

\end{document}